\documentclass[%
 reprint,
superscriptaddress,
 amsmath,amssymb,
 aps,
]{revtex4-1}

\usepackage{booktabs}
\usepackage{multirow}

\usepackage{subcaption}

\usepackage{bm} 
\usepackage{graphicx}
\usepackage{dcolumn}
\usepackage{bm}
\usepackage{hyperref}

\begin{document}

\preprint{APS/123-QED}

\title{Shadow constraints of charged black hole with scalar hair and gravitational waves from extreme mass ratio inspirals}
\author{Lai Zhao}

\author{Meirong Tang}

\author{Zhaoyi Xu}%
\email{zyxu@gzu.edu.cn(Corresponding author)}
\affiliation{%
 College of Physics,Guizhou University,Guiyang,550025,China
}%


\begin{abstract}
Black hole (BH) shadow observations and gravitational wave astronomy have become crucial approaches for exploring BH physics and testing gravitational theories  in extreme environments. This paper investigates the charged black hole with scalar hair (CBH-SH) derived from the Einstein-Maxwell-conformal coupled scalar (EMCS) theory. We first constrain the parameter space $(Q/M, s/M^2)$ of the BH using  the Event Horizon Telescope (EHT) observations of M87* and Sgr A*. The results show that M87* provides stronger constraints on positive scalar hair, constraining the scalar hair $s$ within $0\le s/M^2\le0.4632$ and the charge $Q$ within the range $0\le Q/M\le0.6806$. In contrast, Sgr A* imposes tighter constraints on negative scalar hair. When $Q$ approaches zero, $s$ is constrained within the range $0\geq s/M^2\geq-0.0277$. Overall, EHT observations can provide constraints at most on the order of $\mathcal{O}\left({10}^{-1}\right)$. Subsequently, we construct extreme mass ratio inspiral (EMRI) systems and calculate their gravitational waves to assess the detection capability of the LISA detector for these BHs. The results indicate that for central BHs of $M={10}^6M_\odot$, LISA is expected to detect scalar hair $s/M^2$ at the $\mathcal{O}\left({10}^{-4}\right)$ level and charge $Q/M$ at the $\mathcal{O}\left({10}^{-2}\right)$ level, with detection sensitivity far exceeding the current EHT capabilities. This demonstrates the immense potential of EMRI gravitational wave observations in testing EMCS theory.

\begin{description}
\item[Keywords]
Einstein-Maxwell-conformal coupled scalar; Event Horizon Telescope; EMRI; LISA detector.
\end{description}
\end{abstract}

\maketitle


\section{\label{sec:level1}Introduction}
BHs predicted by general relativity have been confirmed through various astronomical observations, with pivotal evidence such as the inaugural gravitational wave detection from binary BH coalescence in 2015 \cite{LIGOScientific:2016aoc,LIGOScientific:2016sjg}, and the images of the supermassive BH's M87* and Sgr A*, released by the EHT in 2019 and 2022, respectively \cite{EventHorizonTelescope:2019ths,EventHorizonTelescope:2019pgp,EventHorizonTelescope:2019dse,EventHorizonTelescope:2022xqj,EventHorizonTelescope:2022wkp}. These observational achievements confirm BH existence in the universe while simultaneously providing strong evidence for general relativity's validity. While general relativity has been verified in many aspects, it is not the ultimate theory. It faces significant challenges in explaining cosmological phenomena such as the nature of dark matter and the mechanism driving dark energy (see, e.g., \cite{Clifton:2011jh,Ishak:2018his,Fernandes:2021ysi}), spacetime singularity problems \cite{Penrose:1964wq,Hawking:1970zqf,Hawking:1973uf}, the BH information paradox (see, e.g., \cite{Mathur:2009hf,Hawking:1976ra,Hawking:1974rv,Hawking:1975vcx}), and its fundamental incompatibility with quantum theory (see, e.g., \cite{Addazi:2021xuf,Ambjorn:2001cv,Rovelli:2004tv} ). This has motivated the physics community to actively explore theoretical models that go beyond or modify general relativity, as discussed in \cite{Rovelli:2004tv,Ashtekar:2004eh,Clifton:2011jh,Hu:2007nk,Doneva:2017bvd,Antoniou:2017acq,Herdeiro:2018wub,Horndeski:1974wa,Horava:2009if} and references therein.

Among various modified gravity theories, those incorporating scalar fields have received considerable attention. This is because scalar fields not only are crucial in the Standard Model of particle physics (Higgs mechanism) and early universe evolution (see, e.g., \cite{ATLAS:2012yve,Faraoni:2004pi}), but also are considered potential viable models for dark matter or dark energy (see, e.g., \cite{Matos:2023usa,Aybas:2021nvn,Garcia-Arroyo:2024tqq}). In the classical gravity framework, the no-hair theorem asserts that BH properties  are characterized solely by mass $M$, spin $J$, and charge $Q$ \cite{Ruffini:1971bza,Hawking:1971vc,Israel:1967wq}. Therefore, introducing scalar fields in BHs may help test the no-hair theorem. Generally, when scalar fields couple minimally to gravity, boundary conditions prevent the construction of scalar hair BH solutions that violate the no-hair theorem; however, counterexamples exist in the case of non-minimal coupling. 
For instance, in Einstein's conformal coupling scalar theory, the authors in the literature  \cite{Zou:2019ays, Bekenstein:1974sf} constructed an exact BH solution with scalar hair (the Bocharova–Bronnikov–Melnikov–Bekenstein BH), however, the scalar field of this solution exhibits divergence at the event horizon. 
To address this issue, Martínez et al. hid the scalar field singularity within the event horizon by introducing a cosmological constant; however, they only obtained non-planar solutions \cite{Martinez:2002ru,Martinez:2005di}.
 To obtain planar solutions, researchers further introduced the Maxwell field and extended the theory to EMCS theory \cite{Martinez:2005di}. Under this theory, Astorino constructed a charged BH solution with conformal coupled scalar hair \cite{Astorino:2013sfa}.In this paper, we refer to this BH solution as a CBH-SH. The physical properties of this BH have been studied in various aspects, including quasinormal modes \cite{Chowdhury:2018pre} , photon rings and shadows \cite{Myung:2024pob,Khodadi:2020jij}, gravitational lensing \cite{QiQi:2023nex}, and periodic orbits \cite{QiQi:2024dwc}.

In recent years, astronomical observations have achieved breakthrough progress, providing more possibilities for exploring BH properties. On one hand, EHT imaging of supermassive BHs M87* and Sgr A* \cite{EventHorizonTelescope:2019ths,EventHorizonTelescope:2019pgp,EventHorizonTelescope:2019dse,EventHorizonTelescope:2022xqj,EventHorizonTelescope:2022wkp} has provided direct observational evidence for the applicability of general relativity in extreme gravitational regimes. However, uncertainties in observational data also leave room for exploring other modified gravity theories. Therefore, using EHT observational data of M87* and Sgr A*, we can constrain the parameter space of BH models in modified gravity, thereby assessing the plausibility of these BHs in the cosmic environment. On the other hand, as gravitational wave astronomy enters a phase of rapid development,
upcoming spaceborne gravitational wave observatories (including LISA \cite{LISA:2017pwj}, TianQin \cite{TianQin:2015yph,TianQin:2020hid}, and Taiji \cite{Hu:2017mde}) are anticipated to capture gravitational wave signals from extreme-mass-ratio inspiral (EMRI) systems in the millihertz frequency regime \cite{Amaro-Seoane:2007osp,Berry:2019wgg}.
 An EMRI consists of a stellar-mass compact object (CO) (such as a neutron star or Stellar-mass BH) slowly spiraling around a central supermassive BH. The orbital evolution typically lasts for years, with orbital periods reaching thousands or even tens of thousands of cycles. The gravitational wave signals emitted throughout the inspiral process carry rich information about the spacetime structure of the BH. Therefore, EMRI have been applied in many aspects, such as detecting quantum effects (e.g., see \cite{Fu:2024cfk,Yang:2024lmj,Zi:2024jla} etc.), detecting dark matter (e.g., see \cite{Zhang:2024hrq,Zhang:2024ugv,Duque:2023seg,Dai:2023cft} etc.), constraining parameters of other gravitational theories (e.g., see \cite{Zhang:2024csc,Yang:2024cnd,Tan:2024utr,Zhao:2025sck} etc.), testing the no-hair theorem and testing general relativity (e.g., see \cite{Zi:2023omh,Zhao:2024exh,Meng:2024cnq,Qiao:2024gfb,Kumar:2024utz,Zi:2021pdp,Rodriguez:2011aa,Datta:2019euh} etc.).

Based on this background, this paper first uses EHT observational data of supermassive BHs M87* and Sgr A* to constrain the parameters of the charged BH model with scalar hair, thereby assessing the validity of the  EMCS theory in the cosmic environment. Building on this foundation, we construct EMRI systems with the CBH-SH as the central supermassive BH and analyze the effects of different parameters on gravitational waveforms. By calculating the mismatch between waveforms of this BH model and those of Schwarzschild or Reissner-Nordström black holes (RNBH), combined with the expected sensitivity of the LISA detector, we quantitatively determine the parameter magnitudes at which future space-based gravitational wave detectors can identify this BH model. Through comparison with the constraint magnitudes from EHT observational data, we demonstrate the immense potential of EMRI in testing EMCS theory.

The paper structure proceeds as follows: In Section \ref{sec:level2}, we briefly introduce the charged BH with scalar hair in the framework of EMCS theory and systematically derive its geodesic equations. In Section \ref{sec:level3}, we constrain the parameter space of the CBH-SH using EHT observational data. In Section \ref{sec:level4}, by constructing EMRI systems, we analyze the effects of different parameters on waveforms. Additionally, combining the expected sensitivity of the LISA detector, we provide the lower limit magnitudes of parameters that can be effectively identified using the LISA detector.In the last section, the summary and discussion are presented.
 All theoretical derivations in this paper adopt natural units (G=c=1), with appropriate physical dimensions restored during numerical evolution.

\section{\label{sec:level2}Charged Black Hole with Scalar Hair and Geodesics}
In this section, we briefly review the BH solution arising from the interaction between Einstein-Maxwell theory and conformally coupled scalar field, along with the relevant knowledge of geodesics.

The CBH-SH represents an exact solution of EMCS theory. Its existence violates the classical no-hair theorem, demonstrating that a BH can support non-trivial scalar field configuration. The action of this theory consists of the Einstein-Maxwell term coupled with a conformally coupled scalar field term \cite{Martinez:2005di}
\begin{align}
S=&\frac{1}{16\pi G}\int d^4x\sqrt{-g}\left[R-F_{\mu\nu}F^{\mu\nu} \right.\nonumber\\ 
&\left.-8\pi G\left(\nabla_\mu\psi\nabla^\mu\psi+\frac{R}{6}\psi^2\right)\right].
\label{eq1}
\end{align}
Evidently, the above expression describes a coupled system of gravitational, electromagnetic, and scalar fields. Here, $R$ denotes the Ricci scalar, $F_{\mu\nu}$ represents the electromagnetic field strength tensor, $\psi$ is the scalar field, and $\frac{R}{6}\psi^2$ represents the conformal coupling term between the scalar field and curvature.

Using the variational principle, we can derive the field equations for action \eqref{eq1} as
\begin{equation}
R_{\mu\nu}-\frac{1}{2}g_{\mu\nu}R=2T_{\mu\nu}^M+T_{\mu\nu}^\psi,
\label{eq2}
\end{equation}

\begin{equation}
\nabla_\mu F^{\mu\nu}=0,
\label{eq3}
\end{equation}

\begin{equation}
\nabla^2\psi-\frac{R}{6}\psi=0.
\label{eq4}
\end{equation}
In the above, the energy-momentum tensor (EMT) $T_{\mu\nu}^M$ for the electromagnetic field is
\begin{equation}
T_{\mu\nu}^M=\left(F_{\mu\alpha}F_\nu^\alpha-\frac{1}{4}g_{\mu\nu}F_{\alpha\beta}F^{\alpha\beta}\right),
\label{eq5}
\end{equation}
while the scalar field EMT $T_{\mu\nu}^\psi$ is
\begin{align}
T_{\mu\nu}^\psi=&\frac{8\pi G}{6}\bigg[\psi^2\left(R_{\mu\nu}-\frac{R}{2}g_{\mu\nu}\right)+g_{\mu\nu}\nabla^2\left(\psi^2\right)\nonumber\\
&-\nabla_\mu\nabla_\nu\left(\psi^2\right)+6\nabla_\mu\psi\nabla_\nu\psi-3\left(\nabla\psi\right)^2g_{\mu\nu}\bigg].
\label{eq6}
\end{align}

For static spherically symmetric spacetime configurations, the above equations can yield a CBH-SH, which takes the form \cite{Astorino:2013sfa}
\begin{equation}
ds^2=-f\left(r\right)dt^2+f^{-1}\left(r\right)dr^2+r^2\left(d\theta^2+\sin^2{\theta}d\phi^2\right),
\label{eq7}
\end{equation}
where
\begin{equation}
f\left(r\right)=1-\frac{2M}{r}+\frac{Q^2+s}{r^2}.
\label{eq8}
\end{equation}
Here $Q$ is the charge parameter and $s$ is the scalar hair parameter. Furthermore, the corresponding matter fields are
\begin{equation}
\psi=\pm\sqrt{\frac{6}{8\pi G}\sqrt{\frac{s}{s+Q^2}}},
\label{eq9}
\end{equation}
and
\begin{equation}
A_\mu=-\frac{Q}{r}\delta_\mu^t.
\label{eq10}
\end{equation}
Notably, the scalar field of this BH is regular.

\begin{figure}[]
\includegraphics[width=0.45 \textwidth]{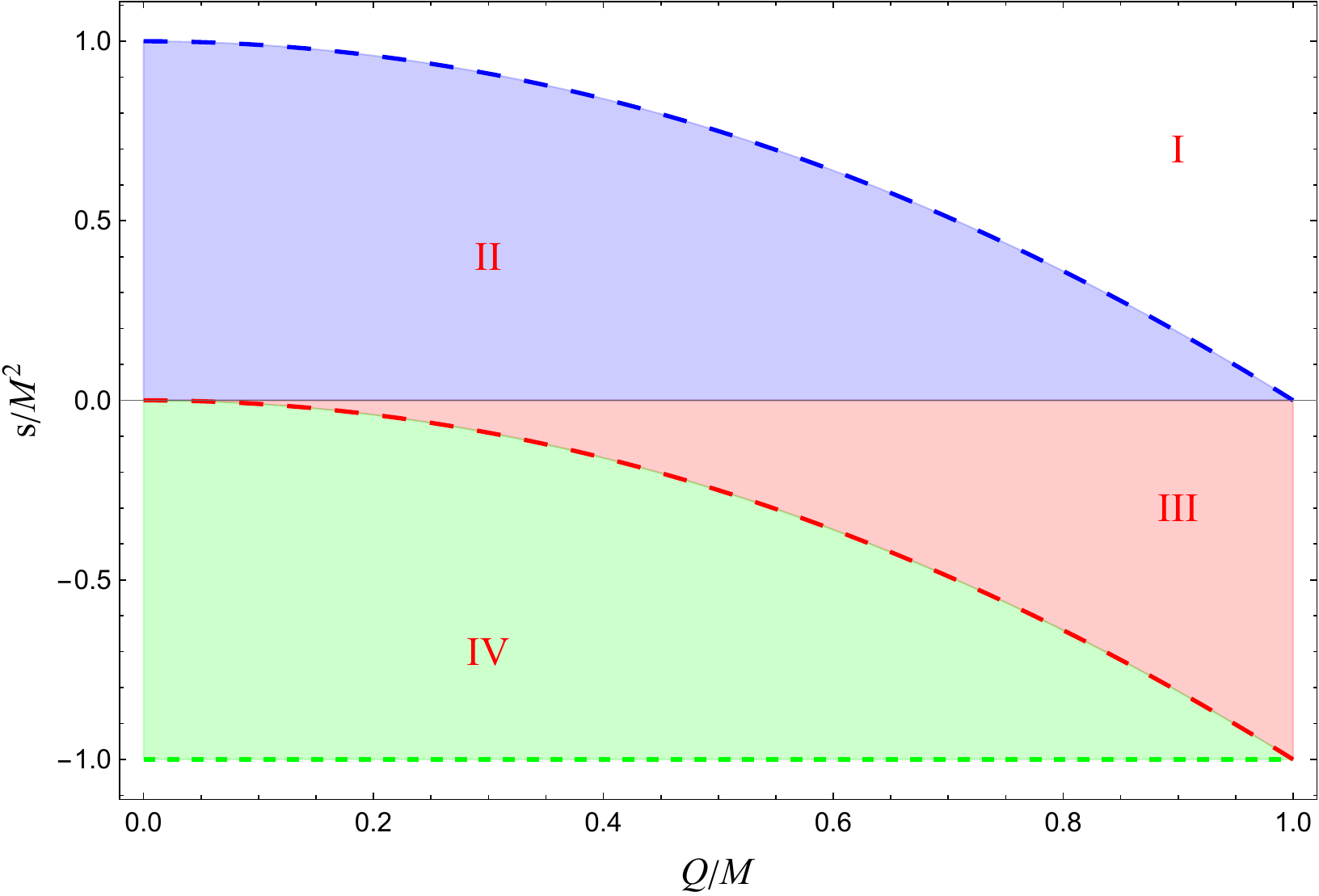}
\caption{
Spatiotemporal structure distribution diagram under different parameter conditions. Region I represents spacetime configurations without BH; regions II, III, and IV characterize situations with BH. Notably, the scalar field exhibits imaginary values in region III, therefore it is not analyzed in this study.}
\label{fig1}
\end{figure}

Obviously, the spacetime structure is influenced by the choice of parameter configuration. The spacetime structure can be determined by $f\left(r\right)=0$, i.e., $r_\pm=M\pm\sqrt{M^2-Q^2-s}$. As shown in Figure \ref{fig1}, the parameter space can be divided into four regions, with boundary characteristics as follows: the blue dashed line represents $M^2=Q^2+s$, corresponding to the critical condition for a extremal BH(where only one event horizon exists); the red dashed line represents $s=-Q^2$, corresponding to the typical Schwarzschild BH structure. According to the parameter values, the physical meaning of each region is: Region I ($M^2<Q^2+s$) represents the parameter space where no event horizon exists, manifesting as naked singularities; Region II ($M^2-Q^2\geq s\geq0$) corresponds to the case with double event horizons, exhibiting a Reissner-Nordström-like BH spacetime structure; in Region III ($-Q^2\le s<0$), although mathematical solutions exist, the scalar field takes imaginary values, therefore it is not within the scope of this study; in Region IV ($s<-Q^2$), scalar field effects on the metric manifest as a $-\frac{1}{r^2}$ term, which differs fundamentally from the behavior of traditional RNBH, hence it is called the ``mutated RNBH".

Considering a test particle moving around the spacetime metric \eqref{eq8}, its Lagrangian can be written as 
\begin{align}
\mathcal{L}&=\frac{1}{2}g_{\mu\nu}\frac{dx^\mu}{d\lambda}\frac{dx^\nu}{d\lambda}\nonumber\\
&=\frac{1}{2}\left[-f\left(r\right)\dot{t}^2+\frac{1}{f\left(r\right)}\dot{r}^2+r^2\dot{\theta}^2+r^2\sin^2{\theta}\dot{\phi}^2\right]=\frac{\epsilon}{2}.
\label{eq11}
\end{align}
Here $\dot{x}^\mu=\frac{dx^\mu}{d\lambda}$ represents differentiation with respect to the affine parameter $\lambda$. Where $\epsilon=-1$ corresponds to timelike particles (massive test particles), and $\epsilon=0$ corresponds to lightlike particles (photons). According to the Lagrangian, the corresponding generalized momentum can be defined as 
\begin{equation}
p_\mu=\frac{\partial\mathcal{L}}{\partial\dot{x}^\mu}.
\label{eq12}
\end{equation}
Combining equations (\ref{eq11}) and (\ref{eq12}), we can obtain the components as
\begin{equation}
p_t=\frac{\partial\mathcal{L}}{\partial\dot{t}}=-f\left(r\right)\dot{t}=-E,
\label{eq13}
\end{equation}

\begin{equation}
p_r=\frac{\partial\mathcal{L}}{\partial\dot{r}}=\frac{\dot{r}}{f\left(r\right)},
\label{eq14}
\end{equation}

\begin{equation}
p_\theta=\frac{\partial\mathcal{L}}{\partial\dot{\theta}}=r^2\dot{\theta},
\label{eq15}
\end{equation}

\begin{equation}
p_\phi=\frac{\partial\mathcal{L}}{\partial\dot{\phi}}=r^2\sin^2{\theta}\dot{\phi}=L.
\label{eq16}
\end{equation}
Here the constants $E$ and $L$ are two conserved quantities arising from spacetime symmetries, corresponding to energy and angular momentum.

If test particles move in the equatorial plane($\theta=\frac{\pi}{2}$) and using the normalization condition $g_{\mu\nu}\dot{x^\mu}\dot{x^\nu}=\epsilon$, combined with expressions (\ref{eq13}), (\ref{eq14}), (\ref{eq15}), and (\ref{eq16}), the particle motion equation can be obtained as
\begin{equation}
\dot{r}^2=E^2-V_{\mathrm{eff}}(r).
\label{eq17}
\end{equation}
Here $V_{\mathrm{eff}}(r)$ denotes the effective potential, and the corresponding expression is
\begin{equation}
V_{\mathrm{eff}}(r)=f(r)\left(\frac{L^2}{r^2}-\epsilon\right).
\label{eq18}
\end{equation}
Obviously, when $\epsilon=0$, it corresponds to the effective potential governing null geodesics, and when $\epsilon=-1$, it corresponds to the effective potential governing timelike geodesics.

For the subsequent discussion, we will on the one hand, analyze the shadow characteristics of the CBH-SH based on null geodesics ($\epsilon=0$) and constrain the parameter space via EHT observational data; on the other hand, we will study the dynamical characteristics and gravitational wave radiation of EMRI systems based on timelike geodesics ($\epsilon=-1$), and impose further constraints on the theoretical parameter space by incorporating LISA's detection sensitivity threshold. The multi-dimensional theoretical probing provided by these two independent observational windows is expected to deepen our understanding of the physical properties of the CBH-SH.

\section{\label{sec:level3}Constraints on Parameter Space from EHT Observations}

The EHT collaboration has successfully published observational results for the supermassive BHs  M87*\cite{EventHorizonTelescope:2019dse,EventHorizonTelescope:2019uob,EventHorizonTelescope:2019jan,EventHorizonTelescope:2019ths,EventHorizonTelescope:2019pgp,EventHorizonTelescope:2019ggy} and Sgr A*  \cite{EventHorizonTelescope:2022wkp,EventHorizonTelescope:2022wok}. These groundbreaking observations revealed the central dark region formed by the BH's strong gravitational field preventing photons from escaping—the BH ``shadow". These observational results provide crucial visual evidence for studying BH physics in the strong gravitational field regime. Furthermore, these observational features also provide important experimental testing methods for general relativity and alternative gravity theories. 
For example, relevant scholars have utilized the characteristics of BH shadows to test modified gravity theories and conduct comparative analyses with Einstein's theoretical framework (see, e.g., \cite{Cunha:2018acu,Perlick:2021aok,Allahyari:2019jqz,Gan:2021pwu,Badia:2021kpk,Meng:2022kjs,Meng:2023wgi,Ali:2024mrt,Yunusov:2024xzu,Afrin:2024khy}  and other related literature); they have tested the BH no-hair theorem by analyzing the geometric morphology of shadows and photon rings (see, e.g., \cite{Gan:2021pwu,Khodadi:2020jij,Yang:2024utv,Glampedakis:2023eek,Khodadi:2021gbc,Broderick:2013rlq,Psaltis:2015uza} and other related literature); and based on these observational data, they have imposed constraints on BH model parameters ( see,e.g., \cite{Allahyari:2019jqz,Meng:2022kjs,Meng:2023wgi,Hou:2021okc,Wu:2025hcu,Tan:2025usr,Molla:2025yoh,Zhao:2024elr} and other related literature).

In this section, we primarily utilize EHT observational data to perform parameter space constraint analysis on the CBH-SH. We treat the CBH-SH as theoretical candidate models for M87* and Sgr A*, and through systematic analysis of EHT observational results, we impose rigorous constraints on the parameter space of such BHs. 
With these observational constraints, we can determine the allowed ranges of the scalar hair parameter and charge parameter, thereby assessing the feasibility of such BHs in astrophysical environments.

For the motion of photons, they follow null geodesics ($\epsilon=0$), where the photon's equation of motion can be described by expression (\ref{eq17}), with the corresponding effective potential being
\begin{equation}
V_{\text{eff}} = \frac{f(r)L^2}{r^2}.
\label{eq19}
\end{equation}

\begin{figure}[]
\includegraphics[width=0.45 \textwidth]{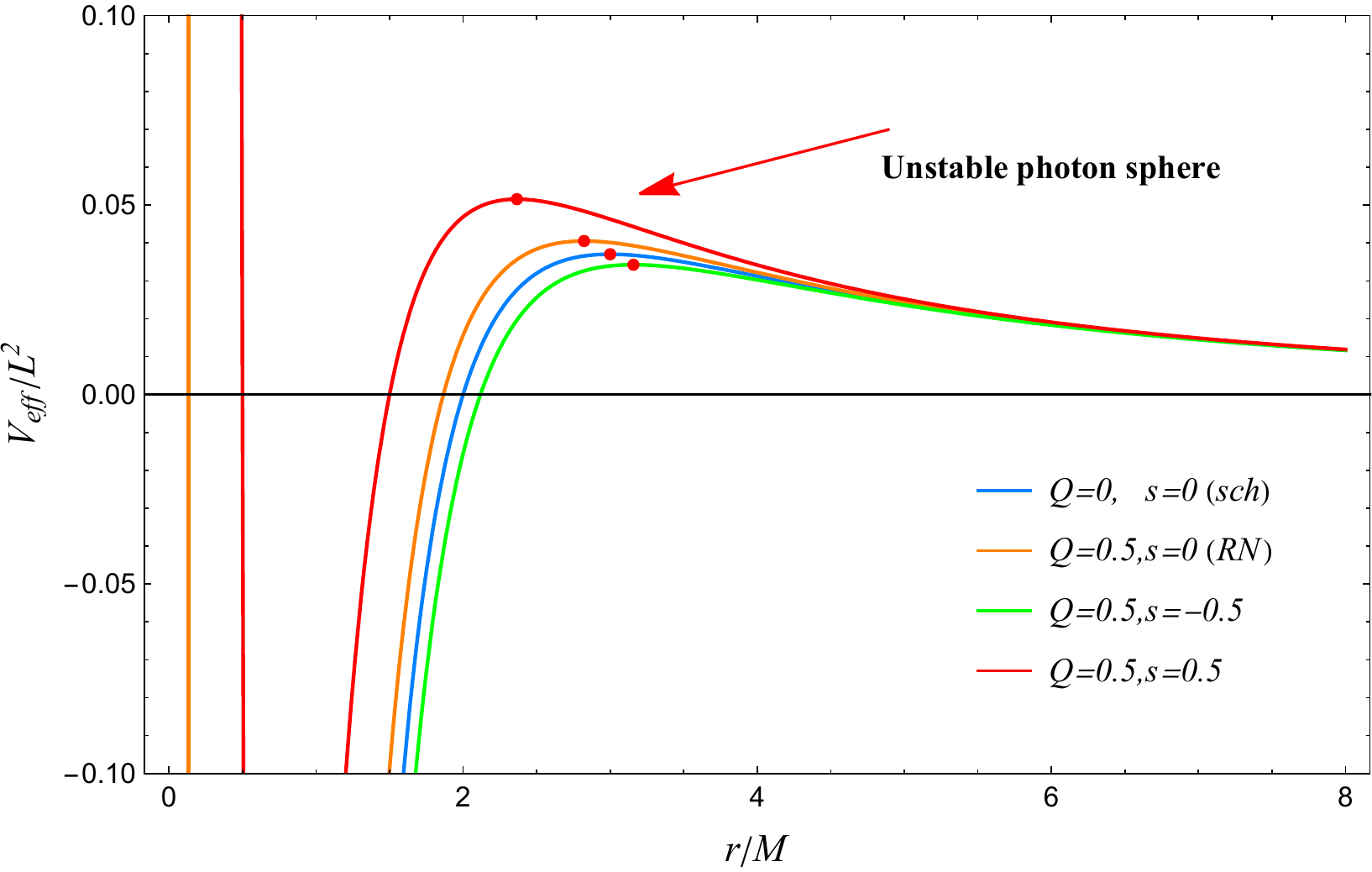}
\caption{
The effective potential of photon motion in different BH models, with red dots marking the corresponding maximum points. Here, the parameters $s$ and $Q$ have been dimensionless, i.e., $s \equiv  s/M^2$, $Q \equiv  Q/M$.
}
\label{fig2}
\end{figure}

For the calculation of the BH shadow radius, we first need to determine the photon sphere radius $r_{\text{ph}}$, which satisfies the conditions
\begin{equation}
V_{\text{eff}} = E^2,\quad\frac{dV_{\text{eff}}}{dr}\bigg|_{r=r_{\text{ph}}} = 0, \;   \frac{d^2V_{\text{eff}}}{dr^2}\bigg|_{r=r_{\text{ph}}} < 0 .
\label{eq20}
\end{equation}
The above conditions are intuitively manifested in the effective potential diagram for photon motion. As shown in Figure \ref{fig2}, we plot the effective potential curves under different BH models, where the red marked points precisely correspond to the positions of unstable photon spheres under various parameter configurations, which corresponds well with condition expression (\ref{eq20}). Clearly, negative scalar hair corresponds to smaller photon sphere radii, while positive scalar hair corresponds to larger photon sphere radii.

If we introduce the impact parameter as $b=\frac{L}{E}$, then combining (\ref{eq19}) and (\ref{eq20}), we derive the critical impact parameter $b_c$ as
\begin{equation}
b_c = \frac{r_{\text{ph}}}{\sqrt{f(r_{\text{ph}})}}.
\label{eq21}
\end{equation}
For an observer at infinity, the theoretically observed BH shadow radius $R_s$ is
\begin{equation}
R_s \approx b_c = \frac{r_{\text{ph}}}{\sqrt{f(r_{\text{ph}})}}.
\label{eq22}
\end{equation}
For the CBH-SH (\ref{eq8}) considered in this paper, the photon sphere radius $r_{\text{ph}}$ can be analytically solved according to formula (\ref{eq20}) as
\begin{equation}
r_{\text{ph}} = \frac{1}{2}\left(3M + \sqrt{9M^2 - 8Q^2 - 8s}\right).
\label{eq23}
\end{equation}

\begin{figure*}[]
\includegraphics[width=1 \textwidth]{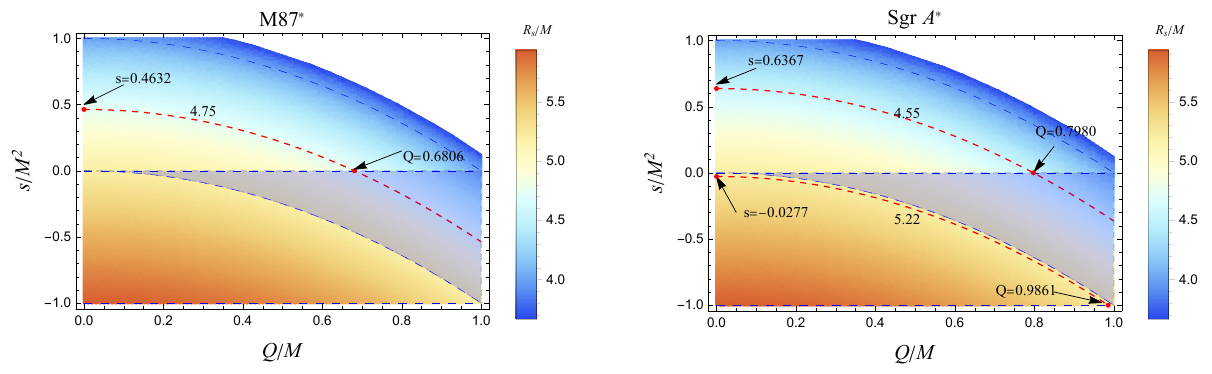}
\caption{EHT observational constraints on the parameter space of the CBH-SH. The left panel shows the constraints on the model parameter space from M87* observational data; the right panel shows the constraints from Sgr A* observational data. The red dashed lines represent the 1$\sigma$ confidence interval boundaries.}
\label{fig3}
\end{figure*}

Combined with the EHT observational results for the supermassive BHs M87* and Sgr A*, the observed shadow radius of M87* is $r_{M87^{\ast}}=(5.5 \pm  0.75)M$ \cite{Allahyari:2019jqz,Bambi:2019tjh}; the observed shadow radius of Sgr A* is $ r_{Sgr\ A^{\ast}}=(4.885\pm 0.335)M$ \cite{Vagnozzi:2022moj}. As shown in Figure \ref{fig3}, we take the CBH-SH as candidates for M87* and Sgr A*, thereby using EHT observational data to preliminarily constrain the parameter space of this BH. The blue dashed lines demarcate the critical thresholds for BH formation, with the enclosed region representing the parameter space where BHs can exist, and the shaded overlapping area is the physically forbidden region (where the scalar field becomes imaginary, which is beyond the scope of this paper's discussion). The relevant discussion of this part has already been presented in the previous section.

Our results show that for a charged BH with positive scalar hair ($s>0$), whether using observational data from M87* or Sgr A*, both can provide good constraints on the parameter space $(s/M^2, Q/M)$ (the region enclosed by red dashed lines represents the 1$\sigma$ confidence interval region). In comparison, using M87* can provide relatively strict constraints on the parameter space $(s/M^2, Q/M)$ (left panel). Specifically, using M87* observational data, the maximum constraint interval for positive scalar hair is $0 \leq s/M^2 \leq 0.4632$, and the maximum constraint interval for charge $Q$ is $0 \leq Q/M \leq 0.6806$. The constraint on the scalar hair parameter becomes more stringent as the charge $Q$ increases, and when reaching the upper limit of $Q$, the constraint on $s$ becomes 0. Using Sgr A* data gives a larger limiting region (right panel). Specifically, using Sgr A* observations can limit the scalar hair to the range $0 \leq s/M^2 \leq 0.6367$ and limit the charge to the range $0 \leq Q/M \leq 0.7980$. For a charged BH with negative scalar hair ($s<0$), M87* does not provide relatively strict constraints, while Sgr A* provides extremely strict constraints (right panel). Specifically, when $Q$ values are close to 0, Sgr A* observational data limits the negative scalar hair to the range $0 \geq s/M^2 \geq -0.0277$.

To quantitatively evaluate the degree of constraint, we define the parameter space overlap area ratio $\eta$ to characterize the constraining capability, i.e., $\eta = \frac{S_{\text{EHT observation region}}}{S_{\text{BH existence region}}} \times 100\%$. The smaller this ratio, the stronger the constraining capability. Our calculations show: For the positive scalar hair case, using M87* data can limit the parameter space within the $\eta = 31.53\%$ region, while using Sgr A* limits it within the $\eta = 49.19\%$ region, indicating that M87* data provides relatively strict constraints on the parameter space. For the negative scalar hair case, the results are dramatically different: M87* data does not impose effective constraints on the parameter space $(s/M^2, Q/M)$ ($\eta = 100\%$), and all the regions we plotted lie within the M87* observation region. However, Sgr A* observational data strictly limits the parameter space within the range of $\eta = 4.12\%$. This result indicates that for a mutated RNBH with negative scalar hair, Sgr A* observational data provides extremely strict constraint conditions, capable of constraining the scalar hair parameter at the $\mathcal{O}(10^{-2})$ level.

Overall, analysis based on EHT observational data shows that the upper limit of parameter space constraints is at the $\mathcal{O}(10^{-1})$ level and below. It should be clearly pointed out that the overlap area ratio $\eta$ here is measured based on the regions plotted in our figure. For the region where scalar hair $s < -M^2$, although not depicted in the figure, this does not significantly affect our analysis conclusions. This is because, as can be clearly seen from Figure \ref{fig3}, when scalar hair $s<-M^2$, it does not significantly affect the parameter constraints.

\section{\label{sec:level4}Gravitational waves from EMRI}

Stellar-mass CO orbiting supermassive BHs form extreme mass ratio binary systems that continuously radiate gravitational waves during their orbital evolution. These gravitational wave signals are among the important detection targets \cite{Amaro-Seoane:2007osp,Berry:2019wgg} for upcoming space-borne gravitational wave observatories such as LISA \cite{LISA:2017pwj}, TianQin \cite{TianQin:2015yph,TianQin:2020hid}, and TaiJi \cite{Hu:2017mde}. Extreme mass ratio binary systems can complete thousands to tens of thousands of orbital cycles before merger, and their waveforms contain information about the spacetime around the central supermassive BH. Therefore, such systems provide unique astrophysical laboratories for testing general relativity, verifying the no-hair theorem of BHs, and constraining BH parameters. Related content can be found in \cite{Fu:2024cfk,Shen:2025svs,Zhang:2024csc,Qiao:2024gfb,Kumar:2024utz,Tan:2024utr,Zi:2021pdp,Datta:2019euh,Zhao:2025sck} and related references.

This section begins by analyzing the characteristics of gravitational waves excited by EMRI systems, followed by the construction of gravitational wave templates using the Augmented Analytic Kludge (AAK) method \cite{Chua:2017ujo,Chua:2015mua,Katz:2021yft}. Finally, based on the expected performance of the LISA detector, we will provide detection thresholds in parameter space and their observational constraints.

\subsection{Flux Evolution and Orbital Evolution}
For binary systems with extreme mass ratios, when a stellar-mass CO orbits a supermassive BH, its motion can be described by timelike geodesics ($\epsilon=-1$). Given that the BH examined in this work is stationary and spherically symmetric, for the sake of simplicity and without compromising generality, we may confine our analysis to motion within the equatorial plane (i.e., $\theta=\frac{\pi}{2}$). Under such simplification, the previously derived equations (\ref{eq13}), (\ref{eq14}), (\ref{eq15}) and (\ref{eq16}) can be rewritten as
\begin{equation}
\left(\frac{dr}{d\tau}\right)^2=\frac{E^2}{m^2}-f\left(r\right)\left(1+\frac{L^2}{m^2r^2}\right),
\label{eq24}
\end{equation}
\begin{equation}
\frac{d\phi}{d\tau}=\frac{L}{mr^2},
\label{eq25}
\end{equation}
\begin{equation}
\frac{dt}{d\tau}=\frac{E}{mf\left(r\right)}.
\label{eq26}
\end{equation}

To better describe the eccentric motion in the equatorial plane, we adopt a parametric representation for the radial coordinate $r$, namely
\begin{equation}
r\left(\psi\right)=\frac{Mp}{1+e\cos\psi},
\label{eq27}
\end{equation}
with $e$ being the orbital eccentricity and $p$ being the semilatus rectum. Under this parametrization, the stellar-mass CO undergoes periodic motion between the pericenter $r_p$ and the apocenter $r_a$, with the two turning points given by
\begin{equation}
r_p=\frac{Mp}{1+e}, \quad r_a=\frac{Mp}{1-e}.
\label{eq28}
\end{equation}
At these two turning points, the radial velocity of the particle becomes zero, i.e., $\frac{dr}{d\tau}=0$. Therefore, by solving equation (\ref{eq24}), we obtain
\begin{align}
E^2&=m^2\left[p\left(2e+p-2\right)+\left(e-1\right)^2\left(Q^2+s\right)\right]\nonumber\\
&\times\frac{\left[p\left(-2e+p-2\right)+\left(e+1\right)^2\left(Q^2+s\right)\right]}{p^2\left[p\left(-e^2+p-3\right)+2\left(e^2+1\right)\left(Q^2+s\right)\right]},
\label{eq29}
\end{align}
\begin{equation}
L^2=\frac{m^2M^2p^2\left(p-Q^2-s\right)}{p\left(-3-e^2+p\right)+2\left(1+e^2\right)\left(Q^2+s\right)}.
\label{eq30}
\end{equation}
When there is no charge and scalar hair, or when $Q^2=s$, the CBH-SH reduces to a Schwarzschild BH, and the above results are consistent with those for a Schwarzschild BH \cite{Hopper:2015icj,Cutler:1994pb}.

For the motion of stellar-mass COs in the equatorial plane, the radial frequency $\Omega_r$ and azimuthal frequency $\Omega_\phi$ can be expressed as
\begin{equation}
\Omega_r=\frac{2\pi}{T_r}, \quad T_r=\int_{0}^{t_0}dt=\int_{0}^{2\pi}\frac{dt}{d\psi}d\psi,
\label{eq31}
\end{equation}
\begin{equation}
\Omega_\phi=\frac{\Delta\phi}{T_r}, \quad \Delta\phi=\int_0^{\phi_0}d\phi=\int_0^{2\pi}\frac{d\phi}{d\psi}d\psi.
\label{eq32}
\end{equation}
Combining equations (\ref{eq25}), (\ref{eq26}) and the parametric expression (\ref{eq27}), we can further obtain
\begin{widetext}
\begin{equation}
\begin{aligned}
\Omega_r=&\frac{(1-e^2)^{3/2}}{Mp^{3/2}}-\frac{3(1-e^2)^{5/2}}{Mp^{5/2}}
+\frac{3(1-e^2)^{5/2}\left[2-6e^2-5\sqrt{1-e^2}+(2+\sqrt{1-e^2})(Q^2+s)\right]}{2Mp^{7/2}}
+O\left(p^{-9/2}\right),
\end{aligned}
\label{eq33}
\end{equation}
\begin{equation}
\begin{aligned}
\Omega_\phi=&\frac{(1-e^2)^{3/2}}{Mp^{3/2}}+\frac{(1-e^2)^{3/2}(6e^2-Q^2-s)}{2Mp^{5/2}}+\frac{(1-e^2)^{3/2}}{8Mp^{7/2}}\bigg[60-18e^2+72e^4-60\sqrt{1-e^2}+60e^2\sqrt{1-e^2}\\
&-12(Q^2+s)-38e^2(Q^2+s)+12\sqrt{1-e^2}(Q^2+s)-12e^2\sqrt{1-e^2}(Q^2+s)-(Q^2+s)^2\bigg]+O\left(p^{-9/2}\right).
\end{aligned}
\label{eq34}
\end{equation}
\end{widetext}

In EMRI, the continuous radiation of gravitational waves leads to gradual loss of the system's energy and angular momentum, a process that dominates the orbital evolution. Here, we use the quadrupole formula for energy flux derived by Peters and Mathews to calculate the dissipated energy and angular momentum \cite{Peters:1964zz,Peters:1963ux}, with the corresponding expressions given by
\begin{equation}
\langle\frac{dE}{dt}\rangle=\frac{1}{5\mu}\left\langle\frac{d^3Q_{ij}}{dt^3}\frac{d^3Q^{ij}}{dt^3}-\frac{1}{3}\frac{d^3Q_{ii}}{dt^3}\frac{d^3Q^{jj}}{dt^3}\right\rangle,
\label{eq35}
\end{equation}
\begin{equation}
\langle\frac{dL_i}{dt}\rangle=\frac{2}{5\mu M}\epsilon_{ijk}\left\langle\frac{d^2Q_{jm}}{dt^2}\frac{d^3Q^{km}}{dt^3}\right\rangle.
\label{eq36}
\end{equation}
Where $\mu=mM/(m+M)$ represents the reduced mass, and $Q_{ij}=\mu x^ix^j$ is the inertia tensor. In the equatorial plane, the corresponding coordinates can be written as $x^i=(r\cos\phi, r\sin\phi, 0)$.

The averaged dissipation rates of energy and angular momentum under the weak-field approximation are
\begin{widetext}
\begin{align}
\left\langle\frac{dE}{dt}\right\rangle=&\frac{(1-e^2)^{3/2}(96+292e^2+37e^4)\mu^2}{15M^2p^5}\nonumber\\
&+\frac{(1-e^2)^{3/2}(e^2(176+450e^2+53e^4)-4(24+104e^2+33e^4)(Q^2+s))\mu^2}{5M^2p^6}+O\left(p^{-7}\right),
\label{eq37}
\end{align}
\begin{align}
\left\langle\frac{dL}{dt}\right\rangle=\frac{4(1-e^2)^{3/2}(8+7e^2)\mu^2}{5Mp^{7/2}}+\frac{2(1-e^2)^{3/2}(2e^2(38+27e^2)-(40+63e^2+2e^4)(Q^2+s))\mu^2}{5Mp^{9/2}}+O\left(p^{-5}\right).
\label{eq38}
\end{align}
\end{widetext}
From the above expressions, we can see that the charge $Q$ and scalar hair parameter $s$ do not appear in the leading-order terms, but rather in the subleading terms. When these parameters vanish or when $s=-Q^2$, the results are consistent with those calculated for a Schwarzschild BH \cite{Hopper:2015icj,Cutler:1994pb}.

For EMRI systems, the orbital period is generally significantly shorter than the evolutionary timescale, thus the adiabatic approximation may be employed, i.e., the system's averaged energy and angular momentum losses are fully radiated away as gravitational waves. The corresponding expressions can be given by the balance laws as
\begin{equation}
\dot{E}_{\mathrm{GW}}=\left\langle\frac{dE}{dt}\right\rangle=-\mu\dot{E},
\label{eq39}
\end{equation}
\begin{equation}
\dot{L}_{\mathrm{GW}}=\left\langle\frac{dL}{dt}\right\rangle=-\mu\dot{L},
\label{eq40}
\end{equation}
where $E$ and $L$ are expressed as functions of $e$ and $p$. Using the chain rule, we obtain
\begin{equation}
-\dot{E}_{\mathrm{GW}}=m\frac{\partial E}{\partial p}\frac{dp}{dt}+m\frac{\partial E}{\partial e}\frac{de}{dt},
\label{eq41}
\end{equation}
\begin{equation}
-\dot{L}_{\mathrm{GW}}=m\frac{\partial L}{\partial p}\frac{dp}{dt}+m\frac{\partial L}{\partial e}\frac{de}{dt}.
\label{eq42}
\end{equation}
Combining equations (\ref{eq39}), (\ref{eq40}), (\ref{eq41}), and (\ref{eq42}), we can derive the orbital evolution equations as
\begin{equation}
m\frac{dp}{dt}=\left[\frac{\partial E}{\partial e}\dot{L}_{\mathrm{GW}}-\frac{\partial L}{\partial e}\dot{E}_{\mathrm{GW}}\right]/\left[\frac{\partial E}{\partial p}\frac{\partial L}{\partial e}-\frac{\partial E}{\partial e}\frac{\partial L}{\partial p}\right], 
\label{eq43}
\end{equation}

\begin{equation}
m\frac{de}{dt}=\left[\frac{\partial L}{\partial p}\dot{E}_{\mathrm{GW}}-\frac{\partial E}{\partial p}\dot{L}_{\mathrm{GW}}\right]/\left[\frac{\partial E}{\partial p}\frac{\partial L}{\partial e}-\frac{\partial E}{\partial e}\frac{\partial L}{\partial p}\right].
\label{eq44}
\end{equation}

During the orbital evolution, the radial phase $\Phi_r$ and azimuthal phase $\Phi_\phi$ also evolve correspondingly, which are associated with the radial frequency $\Omega_r$ and azimuthal frequency $\Omega_\phi$, respectively. The corresponding expressions are

\begin{equation}
\frac{d\Phi_i}{dt}=\Omega_i\left(p\left(t\right),e\left(t\right)\right), \quad i=r,\phi.
\label{eq45}
\end{equation}

To obtain the corresponding waveforms for EMRI, we employ the AAK method \cite{Chua:2017ujo,Chua:2015mua,Katz:2021yft} to construct gravitational wave waveforms. AAK is a fast method for generating EMRI gravitational wave waveforms that combines the efficiency of the AK method \cite{Barack:2003fp} with the accuracy of the NK method \cite{Gair:2005ih,Babak:2006uv}, and has been successfully applied in EMRI gravitational wave signal simulation and preliminary data analysis studies.

Within the transverse-traceless gauge, the two polarization modes may be written in terms of the $n$-th harmonic components \cite{Barack:2003fp}
\begin{align}
h_+ =& \sum_{n} \left\{-\left[1 + \left(\hat{L} \cdot \hat{r}\right)^2\right]\left[a_n \cos 2\gamma - b_n \sin 2\gamma\right]\right\} \nonumber \\
&+ c_n\left[1 - \left(\hat{L} \cdot \hat{n}\right)^2\right],
\label{eq46}
\end{align}

\begin{equation}
h_\times = \sum_{n} 2\left(\hat{L} \cdot \hat{r}\right)\left[b_n \cos 2\gamma + a_n \sin 2\gamma\right].
\label{eq47}
\end{equation}
Where $\gamma = \Phi_\phi - \Phi_r$, $\hat{L}$ is the orbital angular momentum of the stellar-mass CO, and $\hat{r}$ is the direction vector. The complete derivation can be found in reference \cite{Barack:2003fp}. These parameters $(a_n, b_n, c_n)$ are given in terms of Bessel functions of the first kind as \cite{Barack:2003fp}
\begin{align}
a_n =& -nA\left[J_{n-2}(ne) - 2eJ_{n-1}(ne) + \frac{2}{n}J_n(ne) \right.\nonumber\\
&\left.+ 2J_{n+1}(ne) - J_{n+2}(ne)\right]\cos(n\Phi_r),
\label{eq48}
\end{align}
\begin{align}
b_n =& -nA(1-e^2)^{1/2}\left[J_{n-2}(ne) - 2J_n(ne) \right.\nonumber\\&\left.+ J_{n+2}(ne)\right]\sin(n\Phi_r),
\label{eq49}
\end{align}
\begin{equation}
c_n = 2AJ_n(ne)\cos(n\Phi_r).
\label{eq50}
\end{equation}
Here $A = \frac{(M\Omega_\phi)^{2/3}m}{D_L}$, and $D_L$ is the luminosity distance.

In the case of space-deployed gravitational wave detectors such as LISA, the response function is 
\begin{equation}
h^{\mathrm{I,II}}(t) = \frac{\sqrt{3}}{2}\left[h_+(t)F_+^{\mathrm{I,II}}(t) + h_\times(t)F_\times^{\mathrm{I,II}}(t)\right].
\label{eq51}
\end{equation}
Where $F_+^{\mathrm{I,II}}(t)$ and $F_\times^{\mathrm{I,II}}(t)$ are the antenna pattern functions, which characterize the detector's response sensitivity to both gravitational wave polarization components. They are determined by the source's position angles on the celestial sphere $(\theta_s, \phi_s)$ and the detector's azimuthal angles in the orbital plane $(\theta_L, \phi_L)$ \cite{Barack:2003fp,Cutler:1997ta,Apostolatos:1994mx}.

In order to quantitatively evaluate the detection threshold of the LISA detector for the parameters of the CBH-SH, we systematically calculate the waveform mismatch between the CBH-SH and the Schwarzschild BH ($s=0, Q=0$) as well as the RNBH ($s=0, Q\neq0$). The mismatch can be quantified by calculating one minus the overlap, namely
\begin{equation}
\mathcal{M}\left(h_a,h_b\right)=1-\mathcal{O}\left(h_a,h_b\right),
\label{eq52}
\end{equation}
where the normalized overlap $\mathcal{O}\left(h_a,h_b\right)$ is expressed as
\begin{equation}
\mathcal{O}\left(h_a,h_b\right)=\frac{\left\langle h_a\middle| h_b\right\rangle}{\sqrt{\left\langle h_a\middle| h_a\right\rangle\left\langle h_b\middle| h_b\right\rangle}}.
\label{eq53}
\end{equation}
The inner product of two waveforms is defined in the frequency domain as
\begin{equation}
\left\langle h_a\middle| h_b\right\rangle=2\int_{0}^{\infty}{df\frac{h_a^\ast\left(f\right)h_b\left(f\right)+h_a\left(f\right)h_b^\ast\left(f\right)}{S_n\left(f\right)}},
\label{eq54}
\end{equation}
where $h_a\left(f\right)$ and $h_b\left(f\right)$ represent the frequency-domain counterparts of the temporal waveforms $h_a\left(t\right)$ and $h_b\left(t\right)$, correspondingly, and $S_n\left(f\right)$ denotes the power spectral density of noise in the LISA detection system \cite{Maselli:2021men,Robson:2018ifk}. When two signals are identical, the mismatch is zero. For the LISA detector to effectively distinguish between two different gravitational wave signals, their mismatch must satisfy $\mathcal{M}_{th}\geq\frac{D}{2\rho^2}$ \cite{Flanagan:1997kp,Lindblom:2008cm}, where $D=7$ denotes the count of independent variables characterizing the EMRI configuration consisting of the CBH-SH, and $\rho$ is the observed signal-to-noise ratio. The LISA detector design requires a minimum detectable signal-to-noise ratio of 20 \cite{Babak:2017tow}. Under this condition, the critical mismatch threshold for detecting scalar hair $s$ and charge $Q$ is $\mathcal{M}_{th}=0.00875$.

\subsection{Waveform Characteristics and Parameter Evaluation}

In this section, we present the gravitational waveforms under different parameter configurations and their mismatches with various BH models. For quantitative evaluation and comparative analysis, the EMRI system formed by the CBH-SH is constructed with the following parameter configuration: waveform evolution time $t = 1$ year, $p_0 = 15$, $e_0 = 0.1$, $M = 10^6 M_\odot$, $m = 10 M_\odot$, $D_L = 1$ Gpc, $\theta_s = \frac{\pi}{3}$, $\phi_s = \frac{\pi}{2}$, $\theta_L = \frac{\pi}{4}$, $\phi_L = \frac{\pi}{4}$. The scalar hair parameter $s$ and charge parameter $Q$ are specified in the corresponding panels.

\begin{figure*}[ht]
\includegraphics[width=1 \textwidth]{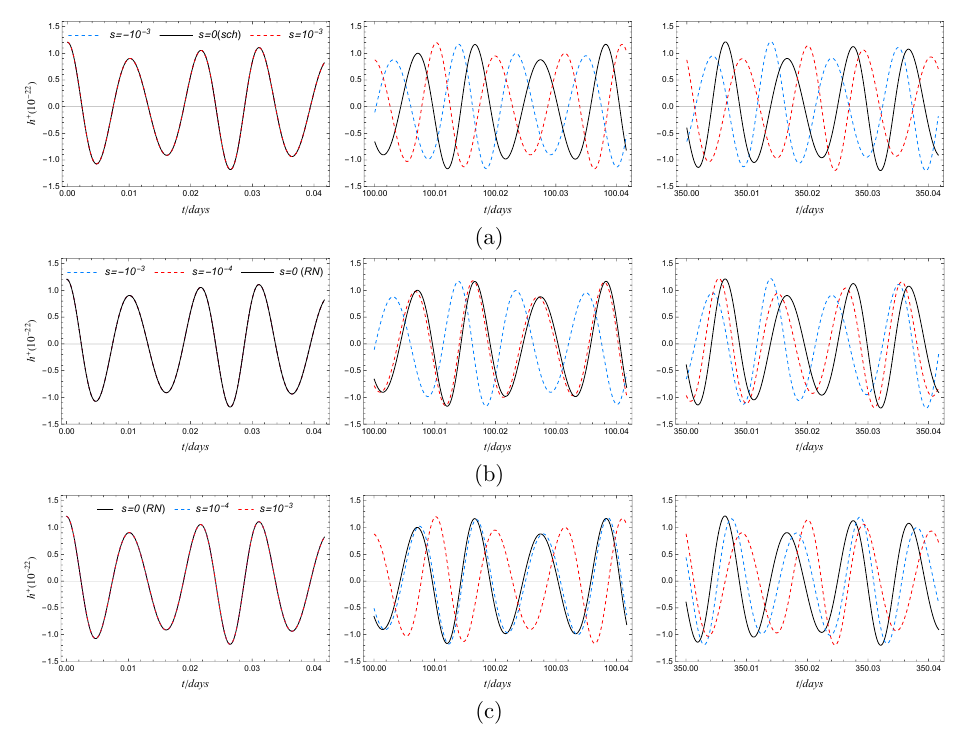}
\caption{
Waveforms for different scalar hair parameters $s$ and charges $Q$. The initial eccentricity and semilatus rectum are $e_0 = 0.1$ and $p_0 = 15$, respectively. (a) Waveforms for scalar hairy BH with $Q = 0$ and $s = \pm 0.001$ (dashed lines) and the Schwarzschild BH with $s = 0$ (solid line); (b) Waveforms for the CBH-SH with $Q = 0.001$ and $s \leq -Q^2$ (dashed lines) and the Schwarzschild BH with $s = -Q^2$ (solid line); (c) Waveforms for the CBH-SH with $Q = 0.001$ and $s > 0$ (dashed lines) and the RNBH with $s = 0$ (solid line).Here, we have already performed dimensionless substitutions, i.e., $s \equiv  s/M^2$ and $Q \equiv  Q/M$.}
\label{fig4}
\end{figure*}

Figure \ref{fig4} illustrates the influence of different scalar hair parameters $s$ and charge $Q$ on the waveform evolution of the system under identical initial conditions. The figure compares the waveform differences between the CBH-SH and both the Schwarzschild BH and the RNBH. The results demonstrate that during the early stages of evolution, the waveforms from different BH models completely overlap; as time evolves, the minute effects introduced by the scalar hair $s$ or charge $Q$ accumulate throughout the long-term evolution process, leading to progressively significant differences between waveforms. It is clearly evident from the figure that even minimal scalar hair parameters or charge $Q$ ( $ |s| \sim 10^{-3} $ or  $Q \sim 10^{-3}$) produce cumulative effects sufficient to generate distinguishable waveform deviations over long-term evolution. This provides potential opportunities for measuring these parameters through space-based detection techniques and conducting comprehensive studies of the fundamental characteristics of scalar hair.

\begin{figure*}[ht]
\includegraphics[width=1 \textwidth]{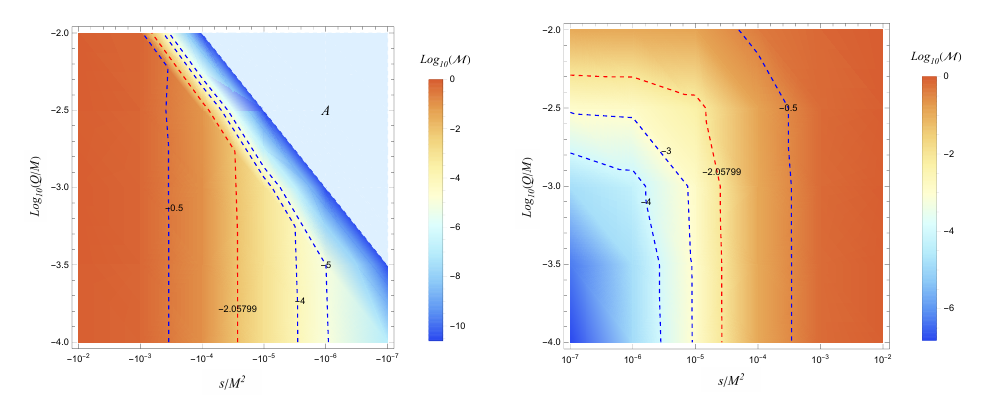}
\caption{
The waveform mismatch between the CBH-SH and the Schwarzschild BH for different parameters ($Q/M$, $s/M^2$). The left panel corresponds to the case of negative scalar hair with $s < Q^2$; the right panel corresponds to the case of positive scalar hair with $s/M^2 > 0$. The initial values of other parameters are set as $M = 10^6 M_\odot$, $p_0 = 15$, and $e_0 = 0.1$.}
\label{fig5}
\end{figure*}

\begin{figure*}[ht]
\includegraphics[width=1 \textwidth]{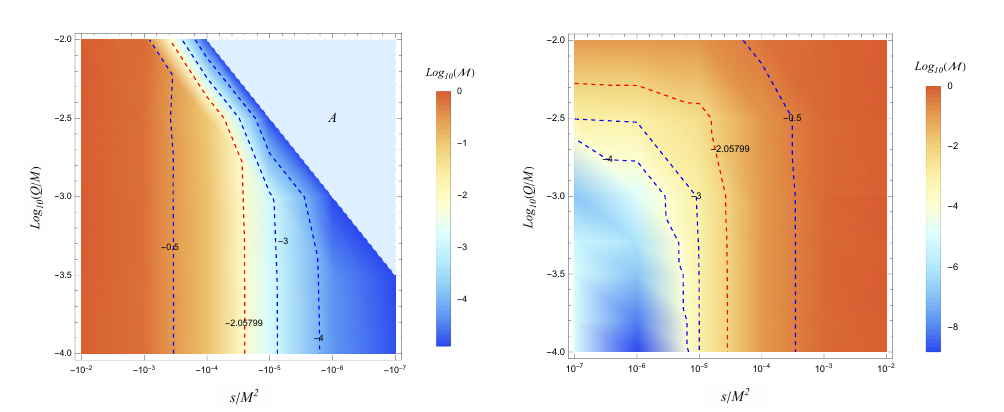}
\caption{
The waveform mismatch between the CBH-SH and the RNBH ($Q/M = 10^{-3}$, $s/M^2 = 0$) for different parameters ($Q/M$, $s/M^2$). The left panel corresponds to the case of negative scalar hair with $s < Q^2$; the right panel corresponds to the case of positive scalar hair with $s/M^2 > 0$. The initial values of other parameters are set as $M = 10^6 M_\odot$, $p_0 = 15$, and $e_0 = 0.1$.}
\label{fig6}
\end{figure*}

\begin{figure*}[ht]
\includegraphics[width=1 \textwidth]{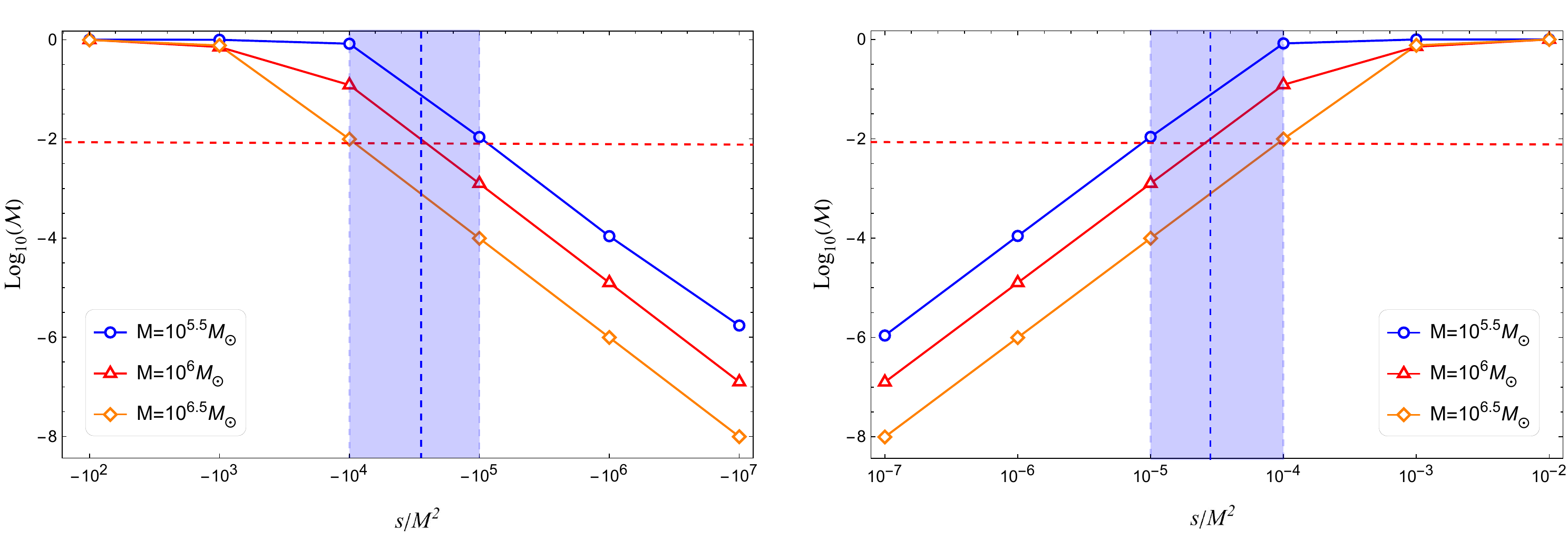}
\caption{
The waveform mismatch between the CBH-SH and the RNBH ($Q/M = 10^{-3}, s/M^2 = 0$) for different central mass values $M$ at a fixed charge parameter $Q/M = 10^{-3}$. The left panel corresponds to the case of negative scalar hair with $s < Q^2$; the right panel corresponds to the case of positive scalar hair with $s > 0$. The initial values are set as $p_0 = 15$ and $e_0 = 0.1$.}
\label{fig7}
\end{figure*}

\begin{figure*}[ht]
\includegraphics[width=1 \textwidth]{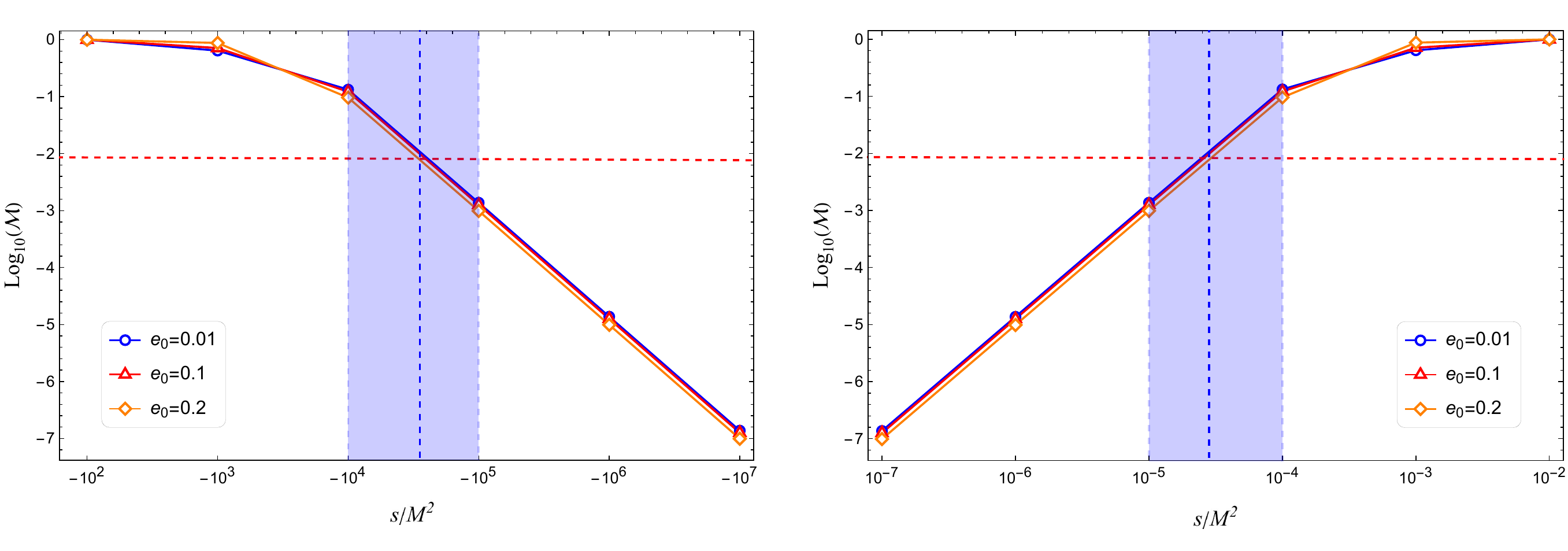}
\caption{
The mismatch between CBH-SH and the RNBH ($Q/M = 10^{-3}, s/M^2 = 0$) for different initial eccentricity values $e_0$ under the condition of a fixed charge parameter $Q/M = 10^{-3}$. The left panel corresponds to the case of negative scalar hair with $s < Q^2$; the right panel corresponds to the case of positive scalar hair with $s/M^2 > 0$. The initial semilatus rectum parameter in the numerical simulation is set as $p_0 = 15$.}
\label{fig8}
\end{figure*}

To quantitatively assess the impact of the parameters $(Q/M, s/M^2)$ of the CBH-SH on gravitational waveforms, we present the waveform mismatch between the CBH-SH and Schwarzschild BH or RNBH for different scalar hair $s$ and charge $Q$ in Figures \ref{fig5}, \ref{fig6}, \ref{fig7}, and \ref{fig8}. The red dashed lines in each figure represent the mismatch threshold $\mathcal{M}_{th} = \log_{10}0.00875 \approx -2.05799$ for the LISA detector to distinguish between two signals, as discussed in the previous section. Furthermore, the data presented in these figures are based on one-year observational evolution.

In Figure \ref{fig5}, we plot the mismatch between the CBH-SH and the Schwarzschild BH for different parameter combinations $(Q/M,s/M^2)$, where the light blue region $A$ represents the physically forbidden zone (in this region $0>s>Q^2$, the corresponding scalar field is complex). Evidently, for the mutated RNBH corresponding to negative scalar hair (left panel), the LISA detector can identify the existence of scalar hair $s/M^2$ at the $\mathcal{O}(10^{-4})$ level. Meanwhile, the detection precision of the charge $Q/M$ is determined by the magnitude of the scalar hair. For instance, at $s/M^2 \sim 10^{-4}$, the charge parameter $Q/M$ can be detected at the $\mathcal{O}(10^{-2})$ level; when the negative scalar hair takes smaller values, the detector's sensitivity to charge detection correspondingly improves. It is worth mentioning that the dark blue region at the edge of region A in the left panel corresponds to minimal mismatch values ($\mathcal{O}(10^{-10})$). This phenomenon occurs because at the edge of this region we have $s \sim Q^2$, which is extremely close to the Schwarzschild BH condition $s = Q^2$, thus resulting in such small mismatch. For the positive scalar hair case, the LISA detector can similarly detect positive scalar hair $s/M^2$ at the $\mathcal{O}(10^{-4})$ level and the charge $Q/M$ at the $\mathcal{O}(10^{-2})$ level.

In Figure \ref{fig6}, we present the mismatch between the CBH-SH and the RNBH ($Q/M=10^{-3}$). For the mutated RNBH case corresponding to negative scalar hair ($s<-Q^2$), the LISA detector can similarly identify the presence of negative scalar hair at the $\mathcal{O}(10^{-4})$ level. The light blue region $A$ in the figure also indicates the physical forbidden zone ($s>-Q^2$), while the dark blue portion at the edge of region A corresponds to mismatch values at the $\mathcal{O}(10^{-5})$ level. This is because in this region, the mutated RNBH is similar to the Schwarzschild BH case ($s=Q^2$), resulting in smaller mismatch values. For the positive scalar hair case, the mismatch results are similar to those analyzed in Figure \ref{fig5}, where the LISA detector can also identify the presence of positive scalar hair $s/M^2$ and charge $Q/M$ at the $\mathcal{O}(10^{-4})$ and $\mathcal{O}(10^{-2})$ levels, respectively.

Furthermore, to investigate the influence of the central BH mass and initial eccentricity on the detection sensitivity of scalar hair $s/M^2$ in EMRI systems, we fix the charge $Q/M=10^{-3}$ and analyze the mismatch distributions between the CBH-SH and the RNBH ($Q/M=10^{-3}$) for different parameter combinations $(M,s)$ and $(e_0,s)$ in Figure \ref{fig7} and \ref{fig8}. The results indicate that the central BH mass has a significant impact on LISA's ability to identify scalar hair, while the initial eccentricity $e_0$ has a relatively weaker influence.
Specifically, for both the negative scalar hair and positive scalar hair cases, as the central BH mass \( M \) increases, the order of magnitude of the detectable scalar hair \(|s|\) gradually increases (as shown in Figure \ref{fig7}). In other words, it can only be identified when \(|s|\) takes relatively large values.
The variation in orbital eccentricity does not significantly affect the detection sensitivity (as shown in Figure \ref{fig8}). In particular, when the central BH mass $M=10^6M_\odot$ or when the system is in a low orbital eccentricity state, the LISA detector can effectively identify the presence of both positive and negative scalar hair at the $\mathcal{O}(10^{-4})$ level.

In summary, for the case of central BH mass $M=10^6M_\odot$, by analyzing the mismatch between the CBH-SH and either the Schwarzschild BH or the RNBH, we find that both positive and negative scalar hair can be identified at the $\mathcal{O}(10^{-4})$ level, while the charge parameter $Q$ can be identified at the $\mathcal{O}(10^{-2})$ level. These sensitivities also imply that constraints on the parameters $Q/M$ and $s/M^2$ can reach the corresponding orders of magnitude. Compared to the constraint results obtained from BH shadow data, the upper limit constraint on charge $Q/M$ is tightened from $\mathcal{O}(10^{-1})$ to $\mathcal{O}(10^{-2})$, and the upper limit constraint on scalar hair $s/M^2$ is also tightened from $\mathcal{O}(10^{-1})$ or $\mathcal{O}(10^{-2})$ (for negative scalar hair constrained by Sgr A* observational data) to $\mathcal{O}(10^{-4})$. Therefore, the space-based gravitational wave detectors expected to be operational around 2030, with their significantly enhanced measurement precision, are expected to enhance the potential of EMRI systems in scalar hair detection, thereby providing deeper insights into the fundamental characteristics of scalar fields in strong gravity regions.


\section{\label{sec:6}Discussion and conclusions}

BH shadow observations and gravitational wave astronomy provide crucial pathways for probing BH physics in strong gravitational fields and testing gravitational theories. This paper primarily investigates the CBH-SH obtained from EMCS theory, where the scalar field of this BH is regular. We first constrain its parameters $(Q/M, s/M^2)$ using EHT observational data (M87*, Sgr A*). Subsequently, within these constrained parameter ranges, we construct an EMRI system with this BH as the central object and calculate its emitted gravitational waveforms. Through mismatch analysis, we quantitatively evaluate the characteristic imprints of charge $Q/M$ and scalar hair $s/M^2$ in gravitational wave signals. Finally, we assess the detectability of these imprints by combining with LISA's expected sensitivity and conservatively estimate the parameter magnitudes that can reach LISA's detection threshold.

Our results indicate that: EHT observational data provide effective constraints on the parameter space $(Q/M, s/M^2)$ of the CBH-SH. For the positive scalar hair charged BH $(s>0)$ case, M87* observational data demonstrate stronger constraining power than Sgr A*. Specifically, the former can constrain the scalar hair $s$ within the range $0\le s/M^2\le0.4632$ and the charge $Q$ within $0\le Q/M\le0.6806$. When using the overlap area ratio to describe the constraining capability, the parameter space $(Q,s)$ can be constrained within $\eta=31.53\%$. In comparison, Sgr A* data provide more relaxed constraints on the parameter space, limiting it within $\eta=49.19\%$. For the mutated RNBH case corresponding to negative scalar hair $(s<-Q^2)$, M87* observational data cannot provide effective constraints; however, using Sgr A* observational data can impose a relatively strict constraint on the parameter space, with the overlap area ratio $\eta=4.12\%$. Particularly, when the charge approaches zero, Sgr A* data can constrain the scalar hair parameter within the range $0\geq s/M^2\geq-0.0277$. Overall, the upper limit of parameter space constraints using EHT observational data is on the order of $\mathcal{O}\left(10^{-1}\right)$.

Furthermore, we construct an EMRI system with the CBH-SH as the central supermassive BH and calculate its gravitational waveforms using the AAK approximation method. By calculating the mismatch between the waveforms produced by this BH and those from Schwarzschild or RN black holes, and combining with LISA detector's expected sensitivity, we quantitatively evaluate the detectability of charge $Q/M$ and scalar hair $s/M^2$ in gravitational wave signals. The results indicate that in an EMRI system with central BH mass $M=10^6M_\odot$, regardless of whether the scalar hair $s/M^2$ takes positive or negative values, LISA is expected to identify its existence at the $\mathcal{O}\left(10^{-4}\right)$ level; while the charge $Q/M$ can be identified at the $\mathcal{O}\left(10^{-2}\right)$ level. These detection precisions significantly exceed the constraint levels achievable by EHT observations. Additionally, we explore the influence of the central BH mass $M$ and initial orbital eccentricity $e_0$ of the EMRI system on LISA's detection capability. The results show that the central BH mass $M$ has a significant impact on waveform mismatch, with smaller central BH masses being more favorable for detecting the existence of scalar hair at higher precision. The low initial eccentricity $e_0$ does not significantly affect waveform mismatch. Comprehensively, for an EMRI system with $M=10^6M_\odot$ and $e_0=0.1$, the LISA detector is expected to identify scalar hair imprints at the $\mathcal{O}\left(10^{-4}\right)$ level and charge parameters at the $\mathcal{O}\left(10^{-2}\right)$ level. This indicates that once future space-based gravitational wave detectors become operational, EMRI systems are expected to become powerful probes for testing and constraining scalar hair characteristics, thereby deepening our understanding of scalar gravitational theories.

Compared to BH shadow observations, EMRI systems, due to their gravitational wave signals lasting for years, allow the minute effects of charge $Q/M$ and scalar hair $s/M^2$ on orbital evolution to accumulate over long-term waveform observations, thus providing the potential to detect these parameter effects at the $\mathcal{O}\left(10^{-2}\right)$ and $\mathcal{O}\left(10^{-4}\right)$ levels, respectively. It is worth noting that this study is based on a static spherically symmetric BH model, and the gravitational waveforms are obtained using the AAK approximation method, therefore only preliminary parameter estimates can be provided in this paper. In the future, extending such analyses to rotating BH backgrounds that more closely match astrophysical observations will undoubtedly be more physically meaningful. 
Furthermore, adopting black hole perturbation theory or incorporating higher-order Post-Newtonian terms to construct waveforms, combined with Fisher information matrix for parameter estimation, is expected to yield more accurate and robust constraint ranges. These studies will be the focus of our future work.

\section{acknowledgements}
We acknowledge the anonymous referee for a construc tive report that has significantly improved this paper. 
This work was supported by Guizhou Provincial Basic Research Program (Natural Science)(Grant No.QianKeHeJiChu[2024]Young166), the Special Natural ScienceFundofGuizhouUniversity(GrantNo.X2022133),the National Natural Science Foundation of China (Grant No.12365008) and the Guizhou Provincial Basic Research Program (Natural Science)(Grant No.QianKeHeJiChu-ZK[2024]YiBan027 and QianKeHeJiChuMS[2025]680).


\bibliography{ref}
\bibliographystyle{apsrev4-1}

\end{document}